\documentclass[reprint,amssymb, amsmath, aps, superscriptaddress, showpacs, footinbib,  prb]{revtex4-1}
\usepackage{graphicx}

\newcommand{\eff}{\text{eff}}
\newcommand{\AFM}{\text{AFM}}
\newcommand{\FM}{\text{FM}}

\newcommand{\imp}{\text{imp}}

\begin{document}

\title{Frustrated couplings between alternating spin-$\frac12$ chains in AgVOAsO$_4$}

\author{Alexander A. Tsirlin}
\email{altsirlin@gmail.com}
\affiliation{Max Planck Institute for Chemical Physics of Solids, N\"{o}thnitzer
Str. 40, 01187 Dresden, Germany}

\author{Ramesh Nath}
\affiliation{Max Planck Institute for Chemical Physics of Solids, N\"{o}thnitzer
Str. 40, 01187 Dresden, Germany}
\affiliation{School of Physics, Indian Institute of Science Education and Research,
Trivandrum-695016 Kerala, India}

\author{J\"org Sichelschmidt}
\affiliation{Max Planck Institute for Chemical Physics of Solids, N\"{o}thnitzer
Str. 40, 01187 Dresden, Germany}

\author{Yurii Skourski}
\affiliation{Dresden High Magnetic Field Laboratory, Helmholtz-Zentrum Dresden-Rossendorf, 01314 Dresden, Germany}
\author{Christoph Geibel}
\author{Helge Rosner}
\email{Helge.Rosner@cpfs.mpg.de}
\affiliation{Max Planck Institute for Chemical Physics of Solids, N\"{o}thnitzer
Str. 40, 01187 Dresden, Germany}

\begin{abstract}
We report on the crystal structure and magnetic behavior of the spin-$\frac12$ compound AgVOAsO$_4$. Magnetic susceptibility, high-field magnetization, and electron spin resonance measurements identify AgVOAsO$_4$ as a gapped quantum magnet with a spin gap $\Delta\simeq 13$~K and a saturation field $\mu_0H_s\simeq 48.5$~T. Extensive band structure calculations establish the microscopic magnetic model of spin chains with alternating exchange couplings $J\simeq 40$~K and $J'\simeq 26$~K. However, the precise evaluation of the spin gap emphasizes the role of interchain couplings which are frustrated due to the peculiar crystal structure of the compound. The unusual spin model and the low energy scale of the exchange couplings make AgVOAsO$_4$ a promising candidate for an experimental investigation of Bose-Einstein condensation in high magnetic fields.
\end{abstract}

\maketitle

\section{Introduction}
Quantum spin systems exhibit diverse and unusual phenomena in high magnetic fields.\cite{bec-review} While the bulk of the experimental and theoretical work refers to systems of weakly coupled spin dimers, similar physics is expected in bond-alternating (dimerized) spin-$\frac12$ chains that likewise feature a singlet ground state with a spin gap. The couplings between the dimers or dimerized chains tend to reduce the spin gap, and heavily affect the high-field behavior. For example, Bose-Einstein condensation (BEC) of magnons, which is a general paradigm for field-induced antiferromagnetic (AFM) order in dimer magnets, exhibits an exotic two-dimensional (2D) regime in BaCuSi$_2$O$_6$, where interdimer couplings are partially frustrated.\cite{[{For example: }][{}]sebastian2006,*laflorencie2009} The abundance of intriguing theoretical predictions, especially for gapped frustrated systems,\cite{laflorencie,michaud2010,bicu2po6} motivates the search for low-dimensional spin-gap magnets with unusual and potentially frustrated geometry of the exchange couplings.

Efficient experimental studies require materials with weak exchange couplings that allow exploring the full temperature-vs-field phase diagram up to the saturation field $H_s$. Direct connections between the polyhedra of magnetic atoms often lead to strong exchange couplings on the order of 100~K. Resulting saturation fields well exceed 100~T and remain unfeasible for present-day experimental facilities. In structures with non-magnetic polyanions, such as phosphates, silicates, or borates, the superexchange is dominated by M--O--O--M pathways (here, M is a transition metal). The long pathways require a specific arrangement of atomic orbitals and usually lead to weak exchange couplings on low-dimensional spin lattices. For example, V$^{+4}$ phosphates are known as a playground for studying spin-$\frac12$ frustrated square lattices.\cite{nath2008-2,tsirlin2009,nath2009,skoulatos2009,carretta2009} Further on, symmetry restrictions on magnetic orbitals induce a variety of one-dimensional (1D) spin lattices in vanadium compounds with 2D\cite{kaul2003,tsirlin2008} or even three-dimensional (3D)\cite{garrett1997,yamauchi1999,azuma1999} crystal structures.

In the following, we present an interesting implementation of the quasi-1D spin lattice in the 3D crystal structure of AgVOAsO$_4$. An extensive experimental and computational study identifies this compound as a system of alternating spin chains with frustrated interchain couplings. The saturation field of about 50~T enables experimental access to the full temperature-vs-field phase diagram, and makes AgVOAsO$_4$ an appealing material for high-field studies that could probe the BEC physics. The outline of the paper is as follows. After reporting methodological aspects in Sec.~\ref{method}, we present the crystal structure from high-resolution synchrotron powder diffraction (Sec.~\ref{structure}), the experimental study by magnetization and electron spin resonance (ESR) measurements (Sec.~\ref{experiment}), and the microscopic magnetic model from density functional theory (DFT) band structure calculations (Sec.~\ref{model}). The experimental and computational results are compared in Sec.~\ref{discussion} that also provides an outlook for future experiments.
\begin{table*}
\begin{minipage}{14cm}
\caption{\label{tab:parameters}
Lattice parameters of AgVOAsO$_4$ and refinement residuals $R_I/R_p$. The standard deviations are taken from the Rietveld refinement.
}
\begin{ruledtabular}
\begin{tabular}{cccccc}
  $T$ (K) & $a$ (\r A) & $b$ (\r A) & $c$ (\r A) & $\beta$ (deg) & $R_I/R_p$   \\
     20   & 6.70096(1) & 8.84034(2) & 7.26478(2) & 115.180(1)    & 0.028/0.090 \\
    300   & 6.71576(1) & 8.84882(2) & 7.28481(2) & 115.282(1)    & 0.033/0.068 \\
\end{tabular}
\end{ruledtabular}
\end{minipage}
\end{table*}

\section{Methods}
\label{method}
Powder samples of AgVOAsO$_4$ were prepared by annealing the stoichiometric mixture of Ag$_2$O, VO$_2$, and As$_2$O$_5$ in an evacuated silica tube at 600~$^{\circ}$C for 24 hours. To avoid the hydration of As$_2$O$_5$, the reagents were handled in an Ar-filled glove box. According to x-ray diffraction (XRD) data collected with Huber G670 Guinier camera (CuK$_{\alpha1}$ radiation, $2\theta=3-100^{\circ}$ angle range, image-plate detector), the samples contained the AgVOAsO$_4$ phase and a minor impurity of metallic silver [the weight fraction of Ag is 1.5(1)~\%, as found from the Rietveld refinement]. The slight reduction of silver is likely caused by a ready decomposition of Ag$_2$O at elevated temperatures, especially in vacuum. However, our attempts to prepare AgVOAsO$_4$ via annealing in air failed due to the partial oxidation of V$^{+4}$. Since the metallic silver shows temperature-independent magnetic susceptibility, this impurity does not influence any of the results presented below. Note that the additional annealing of the AgVOAsO$_4$ samples above 600~$^{\circ}$C led to the melting and decomposition of the compound.

High-resolution XRD data for structure refinement were collected at the ID31 beamline of European Synchrotron Radiation Facility (ESRF) with a constant wavelength of about 0.4~\r A. The signal was measured by eight scintillation detectors, each preceded by a Si (111) analyzer crystal, in the angle range $2\theta=1-40^{\circ}$. The powder sample was contained in a thin-walled borosilicate glass capillary with an external diameter of 0.5~mm. To achieve good statistics and to avoid the effects of the preferred orientation, the capillary was spun during the experiment. The program JANA2000 was used for the structure refinement.\cite{jana2000}

For magnetization measurements in fields below 5~T and in the $2-380$~K temperature range, we used the laboratory Quantum Design SQUID magnetometer (MPMS~5~T). At higher fields, the magnetization isotherm was collected at 1.4~K in the Dresden High Magnetic Field Laboratory using a pulsed magnet. Details of the experimental setup are described elsewhere.\cite{high-field} The perfect coincidence of the data collected during the uprise and decrease of the field indicated the lack of irreversible effects upon the magnetization of the sample. 

To simulate magnetic susceptibility and high-field magnetization curves, we used the ALPS simulation package\cite{alps} and the quantum Monte-Carlo directed loop algorithm in the stochastic series expansion representation.\cite{dirloop} The simulations were performed for finite lattices with periodic boundary conditions. The size of the finite lattice was fixed to $L=120$ for 1D systems and $L\times L$ with $L=24$ for 2D systems. The convergence with respect to the cluster size and to the aspect ratio (in the case of 2D clusters) was carefully checked. 

The ESR experiments were performed with a commercial CW spectrometer for temperatures between 10~K and 294~K by using a $^{4}$He flow-type cryostat. As a function of an external, static magnetic field $B$, the resonance shows up as absorbed power $P$ of a transversal magnetic microwave field at a frequency of 9.4~GHz (''X-band'') or 34~GHz (''Q-band''). To improve the signal-to-noise ratio, a lock-in technique was used by modulating the static field which yields the derivative of the power absorption $dP/dB$.

Scalar-relativistic DFT band structure calculations were performed within the Perdew-Wang\cite{perdew-wang} local density approximation (LDA) for the exchange-correlation potential. The pre-optimized basis set comprised atomic-like local orbitals of the FPLO scheme (\texttt{FPLO8.50}).\cite{fplo} The calculations were based on the low-temperature crystal structure refined from the XRD data. The use of the room-temperature structural data led to only minor differences in individual exchange couplings.

\begin{figure*}
\includegraphics{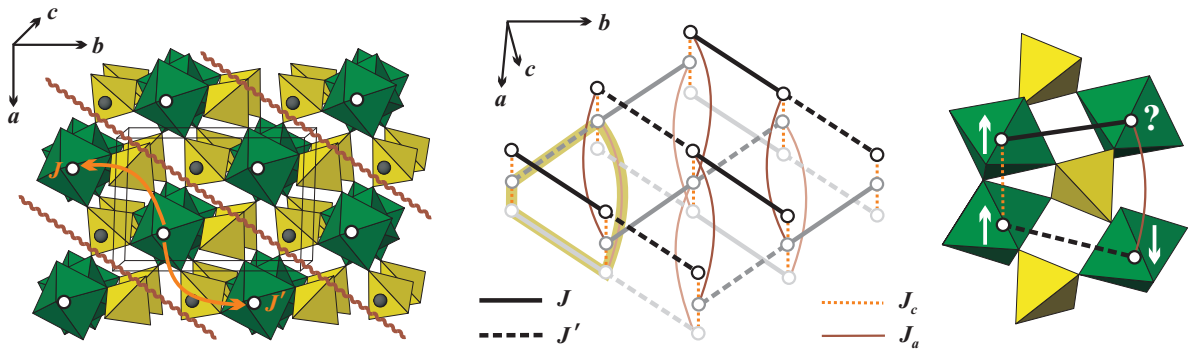}
\caption{\label{fig:structure}
(Color online) Left: crystal structure of AgVOAsO$_4$ with curved lines separating the spin chains. Filled circles show the Ag atoms. Middle: a sketch of the spin model; thick (dark-yellow) line denotes the four-member frustrated unit. Right: the frustrated unit superimposed on the crystal structure. Open circles depict vanadium positions/sites of the spin lattice.
}
\end{figure*}
For the LDA calculations, the crystallographic unit cell and the $k$-mesh of 4096 points (1170 in the irreducible part) were used. The LDA band structure was mapped onto a multi-orbital tight-binding (TB) model, whereas correlations effects in the V $3d$ shell were treated within the Hubbard model. Exchange couplings were derived in the limit of strong correlations (model approach). Alternatively, the correlation effects were treated in a mean-field approximation by local spin density approximation (LSDA)+$U$ calculations for supercells (applying a $k$-mesh of 144 points). Total energies for a number of collinear spin configurations were mapped onto the Heisenberg model (supercell approach). Further details of the computational procedure are given in Sec.~\ref{model}.

\section{Crystal structure}
\label{structure}
The room-temperature x-ray powder pattern of AgVOAsO$_4$ was indexed in a monoclinic unit cell with lattice parameters $a\simeq 6.712$~\r A, $b\simeq 8.849$~\r A, $c\simeq 7.285$~\r A, and $\beta\simeq 115.28^{\circ}$. The reflection conditions $0k0$, $k=2n$ and $h0l$, $l=2n$ identified the $P2_1/c$ space group. The close correspondence to NaVOAsO$_4$ ($P2_1/c$, $a\simeq 6.65$~\r A, $b\simeq 8.71$~\r A, $c\simeq 7.22$~\r A, and $\beta\simeq 115.2^{\circ}$)\cite{navaso5} suggests a similarity of the crystal structures, as further confirmed by the successful Rietveld refinement.\cite{supplement} Atomic displacement parameters (ADPs) of oxygen atoms were constrained (refined as a single parameter) due to the substantial difference in the scattering powers of Ag and~O. 

Tables~\ref{tab:parameters} and~\ref{coordinates} list lattice parameters and refined atomic positions for AgVOAsO$_4$ at 20 and 300~K. Upon cooling, the structure shows an almost uniform shrinkage along all three directions. The atomic positions are nearly unchanged. We also did not observe any non-indexed reflections or reflection splittings that could indicate a lower symmetry or a superstructure.\cite{note1} The noteworthy feature is the relatively high ADP of Ag at 300~K.\cite{note2} At 20~K, the ADP of silver is, however, reduced and becomes comparable to ADP's  of other atoms. Therefore, the high ADP at 300~K should be assigned to thermal vibrations and does not indicate an intrinsic disorder.

\begin{table}
\caption{\label{coordinates}
Atomic positions and isotropic atomic displacement parameters ($U_{\text{iso}}$, in units of $10^{-2}$~\r A$^2$) for AgVOAsO$_4$ at 20~K (first set of lines) and 300~K (second set of lines). All atoms occupy the general $4e$ position of the $P2_1/c$ space group.
}
\begin{ruledtabular}
\begin{tabular}{ccccc}
  Atom & $x/a$     & $y/b$     & $z/c$     & $U_{\text{iso}}$ \\\hline
  Ag   & 0.2418(1) & 0.0846(1) & 0.7204(1) & 0.13(1)          \\
       & 0.2424(1) & 0.0847(1) & 0.7225(1) & 1.49(2)          \\ 
  V    & 0.7441(2) & 0.2376(2) & 0.0300(2) & 0.07(2)          \\
       & 0.7445(2) & 0.2383(1) & 0.0308(2) & 0.39(3)          \\
  As   & 0.2494(2) & 0.0606(1) & 0.2466(1) & 0.11(1)          \\
       & 0.2496(2) & 0.0605(1) & 0.2468(1) & 0.43(2)          \\
  O(1) & 0.8435(8) & 0.0487(5) & 0.6140(6) & 0.00(4)          \\
       & 0.8518(8) & 0.0493(5) & 0.6216(6) & 0.38(4)          \\
  O(2) & 0.0466(8) & 0.1722(5) & 0.0750(6) & 0.00(4)          \\
       & 0.0457(8) & 0.1715(5) & 0.0760(6) & 0.38(4)          \\
  O(3) & 0.7488(9) & 0.1912(4) & 0.2498(7) & 0.00(4)          \\
       & 0.7469(10) & 0.1895(4) & 0.2489(8) & 0.38(4)         \\
  O(4) & 0.6371(8) & 0.0445(5) & 0.8797(6) & 0.00(4)          \\
       & 0.6354(8) & 0.0439(5) & 0.8762(6) & 0.38(4)          \\
  O(5) & 0.4339(8) & 0.1845(5) & 0.4077(6) & 0.00(4)          \\
       & 0.4359(8) & 0.1819(5) & 0.4082(6) & 0.38(4)          \\
\end{tabular}
\end{ruledtabular}
\end{table}

The crystal structure of AgVOAsO$_4$ (Fig.~\ref{fig:structure} and Table~\ref{tab:distances}) is essentially similar to that of NaVOAsO$_4$. Vanadium atoms form distorted octahedra with the short vanadyl bond of about 1.64~\r A. The V--O separations in the equatorial plane of the octahedron amount to $1.99-2.01$~\r A, while the longest V--O distance of 2.14~\r A is found opposite to the vanadyl bond. Arsenic atoms form nearly regular AsO$_4$ tetrahedra with As--O distances of $1.69-1.71$~\r A. Vanadium octahedra share O(3) corners to form chains along the $c$ direction. The AsO$_4$ tetrahedra link the neighboring chains into a 3D framework. 


\begin{table}
\caption{\label{tab:distances}
Interatomic distances (in~\r A) in the AgVOAsO$_4$ structure at 20~K.
}
\begin{ruledtabular}
\begin{tabular}{c@{\hspace{2em}}c@{\hspace{6em}}c@{\hspace{2em}}c}
  Ag--O(1) & 2.463(4) & V--O(1) & 2.010(4) \\
           & 2.536(4) & V--O(2) & 1.996(5) \\
  Ag--O(2) & 2.505(4) & V--O(3) & 1.636(6) \\
  Ag--O(3) & 2.446(4) &         & 2.144(6) \\
  Ag--O(4) & 2.423(5) & V--O(4) & 1.988(4) \\
           & 2.897(4) & V--O(5) & 2.003(5) \\
  Ag--O(5) & 2.487(4) & As--O(1) & 1.698(5) \\
           &          & As--O(2) & 1.712(4) \\
           &          & As--O(4) & 1.697(5) \\
           &          & As--O(5) & 1.693(4) \\
\end{tabular}
\end{ruledtabular}
\end{table}

Magnetic properties of the NaVOAsO$_4$-type compounds have not received a detailed investigation. The magnetic susceptibility of NaVOPO$_4$ was tentatively analyzed within the uniform-chain model,\cite{navpo5-1994} based on the uniform chains of the VO$_6$ octahedra (``structural chains'') running along the $c$ direction. The reference to other V$^{+4}$ compounds, however, does not support this interpretation. A distinctive feature of V$^{+4}$ is the formation of a short bond to oxygen that, in turn, places the unpaired electron of vanadium on the $d_{xy}$ orbital (here, $z$ aligns with the short bond and the structural chain). This arrangement of the magnetic orbital precludes the V--O--V superexchange along the structural chains, and makes the relationship between the crystal structure and the spin lattice rather complex.\cite{garrett1997,kaul2003,nath2008} In Sec.~\ref{model}, we apply extensive DFT calculations in order to find the correct microscopic model. Similar to other V$^{+4}$ compounds, the structural chains do not represent the direction of the leading exchange in AgVOAsO$_4$, although the magnetic behavior is quasi-1D.

\section{Magnetic properties}
\label{experiment}
\subsection{Magnetization}
\label{sec:mag}
\begin{figure}
\includegraphics{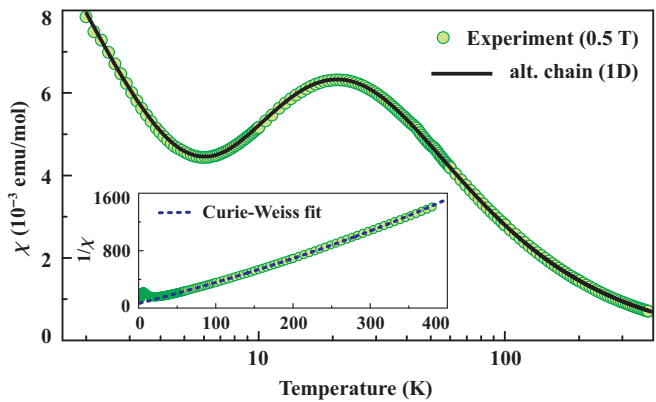}
\caption{\label{fig:chi}
(Color online) Magnetic susceptibility of AgVOAsO$_4$ measured in the applied field of 0.5~T (circles) and the fit with the alternating-chain model [Eq.~\eqref{eq:altchain}] (solid line). The inset shows the Curie-Weiss fit [Eq.~\eqref{eq:cw}, dashed line].
}
\end{figure}

Temperature dependence of the magnetic susceptibility ($\chi$) is shown in Fig.~\ref{fig:chi} and reveals paramagnetic Curie-Weiss behavior above 100~K. At low temperatures, we find a maximum around 20~K followed by an increase in the susceptibility below 6~K. The susceptibility maximum is a characteristic feature of a low-dimensional and/or frustrated system with predominantly antiferromagnetic (AFM) exchange couplings. The low-temperature upturn can be ascribed to the paramagnetism of defects and impurities, as further confirmed by the suppression of this upturn in magnetic field.\cite{supplement} Note that we tried to vary the preparation procedure in order to improve the sample quality, but the low-temperature feature remained nearly unchanged.

The data in the paramagnetic region (above 100~K\cite{note5}) are fitted with the Curie-Weiss law (inset of Fig.~\ref{fig:chi}):
\begin{equation}
  \chi=\chi_0+\dfrac{C}{T+\theta},
\label{eq:cw}\end{equation}
where $\chi_0$ accounts for temperature-independent Van Vleck and diamagnetic contributions, $C$ stands for the Curie constant, and $\theta$ is the Curie-Weiss temperature. We find $\chi_0=-2.0(1)\times 10^{-4}$~emu/mol, $C=0.363(2)$~emu K/mol, and $\theta=20.6(6)$~K. The $C$ value yields the effective magnetic moment of $\mu_{\eff}=1.70(1)~\mu_B$ that closely approaches the spin-only value of 1.73~$\mu_B$ for V$^{+4}$ and corresponds to $g=1.96$ in good agreement with the ESR result of 1.947 (see below). The positive Curie-Weiss temperature $\theta$ identifies leading AFM couplings.

\begin{figure}
\includegraphics{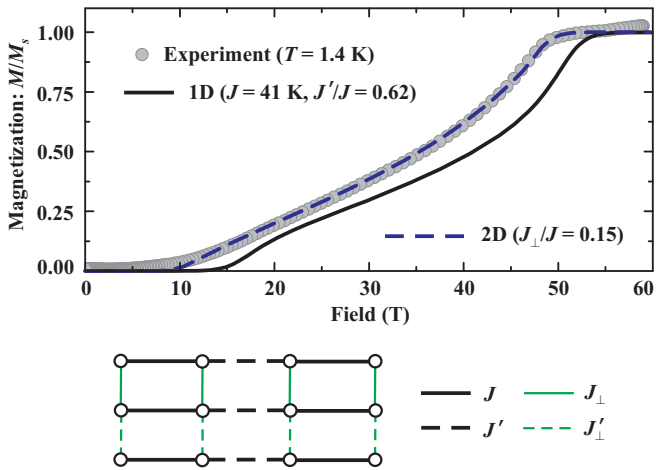}
\caption{\label{fig:mvsh}
(Color online) Top: high-field magnetization curve measured at 1.4~K and corrected for the contribution of paramagnetic impurities;\cite{supplement} solid and dashed lines show the fits of the 1D (isolated alternating chain) and 2D (coupled alternating chains) models, respectively. Experimental values of the magnetization ($M$) are scaled for the saturation magnetization ($M_s$). Bottom: the model of coupled alternating chains.
}
\end{figure}

To fit the susceptibility maximum, a low-dimensional spin model should be applied. Although not obvious at this stage, we use the 1D model of an alternating spin chain, as found from the microscopic analysis (Sec.~\ref{model}). From a phenomenological point of view, this model can be justified by the high-field magnetization curve (Fig.~\ref{fig:mvsh}) that indicates a spin gap in AgVOAsO$_4$. The gapped nature of the excitation spectrum immediately rules out the gapless uniform-chain model. Therefore, the naive structure-based assignment of the spin model (Sec.~\ref{structure} and Ref.~\onlinecite{navpo5-1994}) apparently fails.

Since the data at low temperatures are dominated by the paramagnetic impurity contribution, we fit the susceptibility curve with
\begin{equation}
  \chi=\chi_0+\dfrac{C_{\imp}}{T+\theta_{\imp}}+\chi_{\text{1D}}(J,J'),
\label{eq:altchain}\end{equation}
where $\chi_0$ has the same meaning as in Eq.~\eqref{eq:cw}, the $C_{\imp}/(T+\theta_{\imp})$ term accounts for the low-temperature impurity contribution with $\theta_{\imp}$ being an effective interaction between the impurity spins, and $\chi(J,J')$ is the susceptibility of the alternating spin-$\frac12$ chain with interactions $J$ and $J'$ (at this point, we do not discuss the relation of these couplings to the crystal structure).\cite{johnston2000} Overall,  Eq.~\eqref{eq:altchain} has six variable parameters: $\chi_0,C_{\imp},\theta_{\imp},g,J$, and $J'$. The fit (Fig.~\ref{fig:chi}) yields $\chi_0=-1.9(1)\times 10^{-4}$~emu/mol, $C_{\imp}=0.027(1)$~emu K/mol (7~\% of spin-$\frac12$ paramagnetic impurities), $\theta_{\imp}=1.3(1)$~K, $g=1.87(1)$, $J=41.8(1)$~K, and $J'=25.8(1)$~K. The fitted $g$-value is lower than $\bar g=1.947$ from ESR. Since $g^2$ merely scales with $\chi$, the underestimate is related to the paramagnetic impurity which implies a reduction in the contribution of the AgVOAsO$_4$ phase. Assuming 7~\% of the spin-$\frac12$ impurity, we arrive at $g=1.947\times\sqrt{1-0.07}\simeq 1.88$ in agreement with $g\simeq 1.87$ from the fit. We note that the impurity contribution is not caused by a foreign crystalline phase (this is excluded by synchrotron XRD) and rather originates from AgVOAsO$_4$ itself (e.g., from defects that break the spin chains). Exchange anisotropy of the Dzyaloshinsky-Moriya (DM) type can also be ruled out due to the crystal symmetry (see Sec.~\ref{discussion}).

The low-temperature magnetization curve of AgVOAsO$_4$ (Fig.~\ref{fig:mvsh}) is characteristic of a spin-gap magnet. After subtracting the 7~\% contribution of spin-$\frac12$ paramagnetic impurity approximated by a Brillouin function,\cite{supplement} we find that the magnetization ($M$) is close to zero below $\mu_0H_c\simeq 10$~T, increases above 10~T due to the closing of the spin gap, and saturates at $\mu_0H_s\simeq 48.5$~T. The bend at $H_c$ directly measures the spin gap $\Delta=g\mu_0\mu_BH_c/k_B\simeq 13$~K.\cite{note3}

The comparison to the simulated curve for the alternating spin chain with $J$ and $J'$ from the susceptibility fit underscores two important differences. First, the fitted values of $J$ and $J'$ overestimate the spin gap and the saturation field. Second, the shape of the experimental curve is different from one expected for a purely 1D system: the bend around 20 T is not observed, while the bend at $40-45$~T, preceding the saturation, is notably smeared. Since we simulate the curve at a realistic temperature of $T/J=0.035$ (i.e., 1.4~K), the smearing of the bends cannot be ascribed to thermal fluctuations. The smooth shape of the experimental curve is rather caused by interchain couplings that modify the excitation spectrum of the system.

The 2D model of coupled alternating spin chains ($J-J'-J_{\perp}$) improves the fit of the magnetization curve. We were able to impose either FM, AFM, or both FM and AFM interchain couplings to obtain nearly indistinguishable fits (dashed line in Fig.~\ref{fig:mvsh} and Table~\ref{tab:fits}). The simulations are done in dimensionless units $H/(\bar g\mu_BJ)$, hence both $\bar g$ and $J$ values simply scale the field axis. Therefore, we fixed $\bar g=1.947$ according to ESR and adjusted $J$. We find a higher $J$ for FM interchain couplings, whereas AFM interchain couplings contribute to the saturation field and reduce $J$. However, none of the regimes fits to $\chi(T)$ and $M(H)$ simultaneously. For AFM interchain couplings, the fit of the high-field magnetization underestimates $J$ compared to the susceptibility fit. For FM interchain couplings, the $J$ values from the two fits nearly match, yet the fitted $g$-value for $\chi(T)$ is well below 1.947 (or $\simeq 1.88$ once the impurity contribution is taken into account, see above). The model with FM-and-AFM interchain couplings yields a reasonable $g$-value, although sizable discrepancies in the $J$ value remain (see Table~\ref{tab:fits}).

Presently, we are unable to find a model that will provide a consistent fit of the magnetic susceptibility and high-field magnetization. This situation is drastically different from another alternating-chain-based system Pb$_2$V$_3$O$_9$ (Ref.~\onlinecite{pb2v3o9}) where the introduction of FM interchain couplings readily resolves all the drawbacks of the 1D model. The lack of the consistent fit likely results from the complex and frustrated arrangement of the interchain couplings, as further confirmed by our microscopic analysis in Sec.~\ref{model}.
\begin{table}
\caption{\label{tab:fits}
Fitting of the magnetic susceptibility and high-field magnetization data with the model of coupled alternating spin chains (bottom part of Fig.~\ref{fig:mvsh}): $J_{\perp}=J_{\perp}'>0$ (first line), $J_{\perp}=J_{\perp}'<0$ (second line), and $J_{\perp}=-J_{\perp}'$ (third line). The magnetization curves were fitted with the constant $\bar g=1.947$ from ESR.
}
\begin{ruledtabular}
\begin{tabular}{rcccc}
  $J_{\perp}/J$ & $J'/J$ & $J$ (from $\chi$) & $g$ (from $\chi$) & $J$ (from $M$) \\
      0.15      &  0.60  &  42               & 1.93              & 34             \\
     $-0.27$    &  0.60  &  41               & 1.79              & 40             \\
     $\pm0.15$  &  0.60  &  41               & 1.86              & 37             \\
\end{tabular}
\end{ruledtabular}
\end{table}

\subsection{Electron spin resonance}
\begin{figure}[htbp]
\begin{center}
\includegraphics[width=0.5\textwidth]{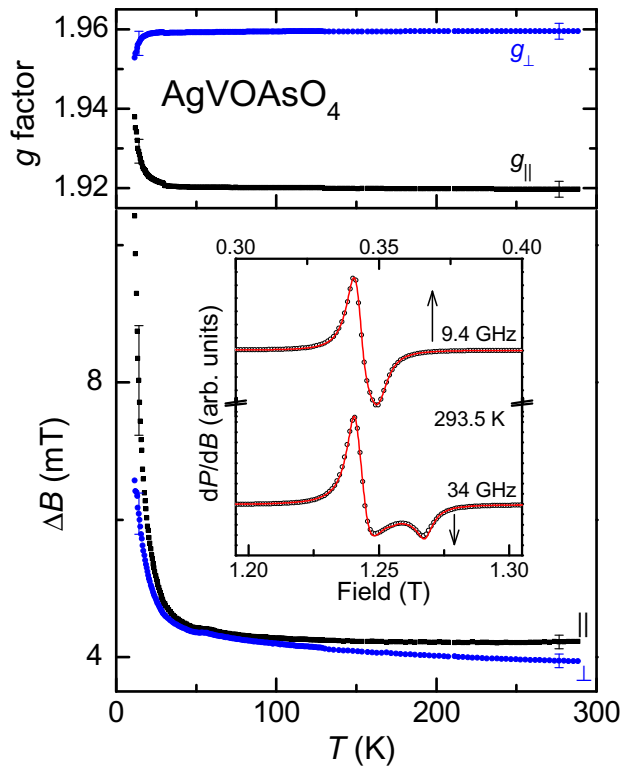}
\caption{
(Color online) Temperature dependence of X-band ESR parameters: $g$ factor (top panel) and ESR linewidth $\Delta B$ (bottom panel). The inset shows the room-temperature spectra (symbols) at two microwave frequencies, X-band (9.4 GHz) and Q-band (34 GHz), together with powder-averaged Lorentzian line fits (solid lines) that provided the axial-symmetric parameters $g_{\|,\perp}$ and $\Delta B_{\|,\perp}$.
}
\label{ESR}
\end{center}
\end{figure}

The ESR experiments on powder samples of \rm{AgVOAsO$_{4}$} yield spectra with a shape typical for powder-averaged Lorentzian lines, see the inset of Fig.~\ref{ESR}. The anisotropy of the $g$ values corresponds to a resonance field anisotropy which is clearly resolved in the Q-band spectra. As shown by the solid fitting lines, the two components of a uniaxial powder fit are sufficient to describe the data (symbols) in a reasonable quality, despite the crystallographic symmetry is monoclinic. This fitting works well in the whole temperature range under investigation. At the X-band frequency and $T=293.5$~K, we obtained $g_{\|}=1.920\pm0.002$ and $g_{\perp}=1.960\pm0.002$ which corresponds to an average $g$ factor \cite{abragam70a} $\bar{g}=\frac{1}{3}(g_{\|}+2g_{\perp})=1.947\pm0.002$. These values are similar to those found for V$^{4+}$ in other compounds with VO$_{6}$ octahedra, see, e.g., Ref.~\onlinecite{ivanshin2003}. The Q-band spectra can be described by the same $g$-values within the experimental error. Note that the difference between $g_{\perp}$ and $g_{\|}$ implies the anisotropy of a single magnetic ion, and does not indicate exchange anisotropy. 

The temperature dependence of the integrated ESR absorption and $\chi(T)$ (Fig.~\ref{fig:chi}) agree to each other (not shown), which confirms the validity of the temperature dependencies of the X-band ESR parameters shown in the main panels of Fig.~\ref{ESR}. Both $g$-values remain constant in a wide temperature range down to 35~K and 22~K for $g_{\|}$ and $g_{\perp}$, respectively. The linewidth $\Delta B_{\|,\perp}(T)$, describing the half-width at half-maximum of the ESR absorption, shows a characteristic behavior as observed in many other low-dimensional spin systems. \cite{ivanshin2003,sichelschmidt2002,moeller2009} It steeply increases below 50~K indicating the onset of spin correlations at low temperatures. Above 50~K, the anisotropy of the linewidth appears to be temperature-dependent: $\Delta B_{\|}(T)$ shows a shallow minimum at around 210~K whereas $\Delta B_{\perp}$ continues to decrease with increasing temperature. The weak anomaly near 50~K is likely related to uncertainties in the fitting procedure. 

\section{Band structure and microscopic model}
\label{model}
\begin{figure}
\includegraphics{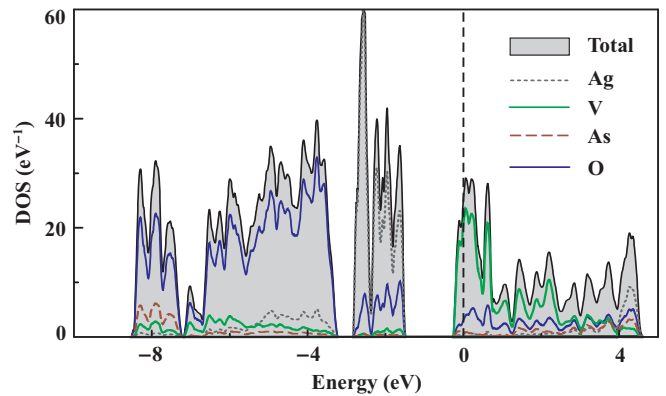}
\caption{\label{fig:dos}
(Color online) LDA density of states for AgVOAsO$_4$. The Fermi level is at zero energy.
}
\end{figure}

To understand the relation between the 1D alternating-spin-chain model and the 3D crystal structure of AgVOAsO$_4$, we perform band structure calculations and evaluate individual exchange couplings. The LDA energy spectrum (Fig.~\ref{fig:dos}) reveals an electronic structure, typical for V$^{+4}$ compounds.\cite{tsirlin2009,nath2008,mazurenko2006} The low-lying valence bands are predominantly formed by O $2p$ states. The bands around $-3$~eV represent the filled $4d$ shell of Ag$^{+1}$ (similar to Ref.~\onlinecite{tsirlin2008}). The bands at the Fermi level have V $3d$ origin and overlap with empty states of Ag ($5s$) and As ($4s$) above 3~eV. A closer examination reveals the $t_{2g}$ and $e_g$ crystal-field levels of V $3d$ shell below and above 1~eV, respectively. The gapless energy spectrum results from the well-known underestimate of correlation effects in LDA.

\begin{figure}
\includegraphics{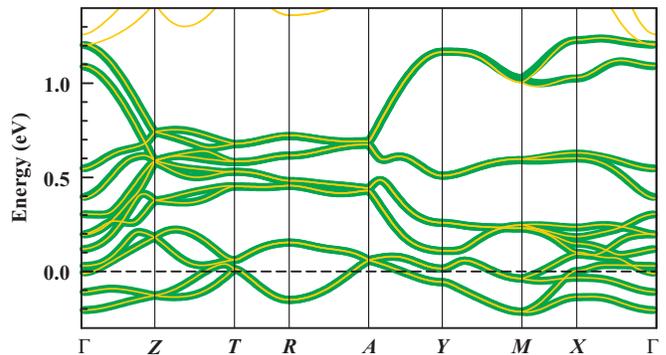}
\caption{\label{fig:bands}
(Color online) LDA band structure of AgVOAsO$_4$ (thin light lines) and the fit with the tight-binding model (thick dark lines). The $k$-space path is defined as follows: $\Gamma(0,0,0)$, $Z(0,0,0.5)$, $T(0.5,0,0.5)$, $R(0.5,0.5,0.5)$, $A(0,0.5,0.5)$, $Y(0,0.5,0)$, $M(0.5,0.5,0)$, $X(0.5,0,0)$ with the coordinates given in units of the respective reciprocal lattice parameters. The Fermi level is at zero energy.
}
\end{figure}

The formation of the short V--O bond typically leads to a crystal-field split of the $t_{2g}$ levels. Then, the half-filled orbital has the $xy$ symmetry and lies in the plane perpendicular to the short bond.\cite{tsirlin2009,mazurenko2006,tsirlin2010} The electronic structure of AgVOAsO$_4$ reminds of this scenario, yet the $t_{2g}$ levels are not fully split. The bands between $-0.2$ and $0.2$~eV indeed have a $d_{xy}$ origin but they are strongly hybridized with the higher-lying $d_{yz}$ and $d_{xz}$ bands (Fig.~\ref{fig:bands}). This can be related to the octahedral coordination of vanadium and to the relatively short \mbox{V--O(3)} bond (2.14~\r A) opposite to the vanadyl bond, similar to Pb$_2$V$_3$O$_9$ (Ref.~\onlinecite{pb2v3o9}).

To include correlation effects and evaluate exchange couplings, we construct a three-orbital tight-binding model (Fig.~\ref{fig:bands}) based on Wannier functions with proper orbital characters.\cite{wannier-fplo} The reduction to the three-orbital model is justified by the low contribution of the $e_g$ states to the exchange.\cite{note4} The orbital energies of 0.05~eV ($xy$), 0.50~eV ($xz$), and 0.60~eV ($yz$) are consistent with our expectations regarding the lowest energy of the $d_{xy}$ level. In the correlated Mott-insulating state (as obtained from LSDA+$U$ calculations), the $d_{xy}$ orbital is half-filled, and the exchange couplings are derived from the Kugel-Khomskii model:\cite{mazurenko2006,kugel1982}
\begin{equation}
  J=\dfrac{4t_{xy}^2}{U_{\eff}}- \sum_{\alpha=yz,xz}\dfrac{4t_{xy\rightarrow\alpha}^2J_{\eff}}{(U_{\eff}+\Delta_{\alpha})(U_{\eff}+\Delta_{\alpha}-J_{\eff})},
\label{eq:exchange}\end{equation}
where $t_{xy}$ and $t_{xy\rightarrow\alpha}$ are the hoppings between the $xy$ states and from the $xy$ (half-filled) to $\alpha$ (empty) states, $U_{\eff}$ and $J_{\eff}$ are the effective on-site Coulomb repulsion and Hund's coupling in V $3d$ bands, respectively, and $\Delta_{\alpha}$ is the crystal-field splitting between the $xy$ and $\alpha$ states. The first term in Eq.~\eqref{eq:exchange} is an AFM coupling due to hoppings between the half-filled orbitals. The second term stands for a FM coupling that arises from hoppings to empty orbitals followed by the Hund's coupling on the vanadium site. Note also that the system is in the strongly correlated regime ($t_i\ll U_{\eff}$).

\begin{table}
\caption{\label{tab:exchanges}
Interatomic distances and exchange integrals (in~K) calculated within the model approach ($U_{\eff}=4$~eV, $J_{\eff}=1$~eV) and the supercell approach (LSDA+$U$, AMF, $U_d=4$~eV, $J_d=1$~eV). 
}
\begin{ruledtabular}
\begin{tabular}{cc@{\hspace{4em}}rrr@{\hspace{4em}}r}
     & Distance & $J^{\AFM}$ & $J^{\FM}$ & $J$       & $J$      \\
     &  (\r A)  & \multicolumn{3}{l}{Model approach} & LSDA+$U$ \\
$J$  &  5.59    &  88        &  0        & 88        & 47       \\
$J'$ &  5.56    &  59        &  $-9$     & 50        & 47       \\
$J_c$ & 3.64    &  0         &  $-4$     & $-4$      & $-9$     \\
$J_a$ & 6.12    &  11        &  0        & 11        & 6        \\
\end{tabular}
\end{ruledtabular}
\end{table}
\begin{table}
\caption{\label{tab:lsda+u}
Exchange integrals (in~K) calculated via the supercell approach with different values of LSDA+$U$ Coulomb repulsion parameter $U_d$ (in~eV) and different DCC schemes, see text for details.
}
\begin{ruledtabular}
\begin{tabular}{cccccc}
  $U_d$ & $J$ & $J'$ & $J_c$ & $J_a$ &     \\
    3   & 62  & 61   & $-2$  & 7     & AMF \\
    4   & 47  & 47   & $-9$  & 6     & AMF \\\smallskip
    5   & 35  & 38   & $-24$ & 5     & AMF \\
    4   & 41  & 36   & $-8$  & 4     & FLL \\
\end{tabular}
\end{ruledtabular}
\end{table}
To derive $J$'s, we take $U_{\eff}=4$~eV and $J_{\eff}=1$~eV, as established for other V$^{+4}$ compounds.\cite{mazurenko2006,tsirlin2010,pb2v3o9} The resulting exchanges are listed in Table~\ref{tab:exchanges}. The leading couplings $J$ and $J'$ are AFM and run along $[110]$ or $[1\bar 10]$. This leads to alternating spin chains that align with different crystallographic directions to form an unusual crossing pattern (middle panel of Fig.~\ref{fig:structure}). The 	couplings between the spin chains, FM $J_c$ (along the structural chains) and AFM $J_a$ (approximately along $[102]$), complete our model to a peculiar 3D spin lattice. Further couplings are below 1~K and can be neglected.

LSDA+$U$ supercell calculations (last column of Table~\ref{tab:exchanges}) confirm the results of the model analysis. We find leading AFM couplings $J$ and $J'$ along with weaker interchain couplings, FM $J_c$ and AFM $J_a$. Note, however, that the alternation ratio can not be evaluated precisely. The model approach yields $\alpha=J'/J=0.57$ in remarkable agreement with the experimental $\alpha=0.60-0.65$, yet the absolute values of $J$ and $J'$ are overestimated by a factor of two (e.g., $J=88$~K in DFT vs. $J\simeq 40$~K in the experiment). LSDA+$U$ better evaluates the absolute values but overestimates the $\alpha$ ratio. 

To get additional insight regarding the influence of computational parameters on individual exchange couplings, we calculated exchange integrals for different values of $U_d$, the on-site Coulomb repulsion parameter of the LSDA+$U$ method (Table~\ref{tab:lsda+u}). We also applied different double-counting correction (DCC) schemes, around-mean-field (AMF) and fully-localized-limit (FLL), since the choice between the two is known to affect exchange couplings in Cu$^{+2}$ compounds.\cite{bicu2po6,cu2v2o7,volborthite} Surprisingly, the exchange couplings in AgVOAsO$_4$ are largely insensitive to the DCC, whereas the change in $U_d$ leads to a typical evolution of $J$'s (see, e.g., Ref.~\onlinecite{nath2008}). The $\alpha\simeq 1$ ratio is obtained for a range of $U_d$ values, hence this result should be taken as an intrinsic drawback of LSDA+$U$ calculations for AgVOAsO$_4$. Overall, we find that both the model and supercell approaches experience certain difficulties in the precise evaluation of the exchange couplings. The combination of different methods allows to derive the complete spin model, and facilitates the microscopic understanding of the material, whereas the extensive analysis of the experimental data (Sec.~\ref{experiment}) justifies our assignment of AgVOAsO$_4$ to the model of coupled alternating spin chains.

\section{Discussion and summary}
\label{discussion}
The experimental results for the magnetic susceptibility and high-field magnetization of AgVOAsO$_4$ can be qualitatively understood within the bond-alternating spin chain model. The alternation of the exchange couplings $J$ and $J'$ opens the spin gap $\Delta\simeq 13$~K, and leads to the likely singlet ground state (no long-range order in zero field). The microscopic origin of $J$ and $J'$ is rationalized by DFT. Inversion centers of the $P2_1/c$ space group relate the neighboring vanadium atoms within the spin chains, thereby DM couplings on the $J$ and $J'$ bonds are forbidden by symmetry. Despite the weak anisotropy of the $g$-tensor, the exchange couplings in AgVOAsO$_4$ are, to a good approximation, isotropic. 

The alternating-chain magnetic behavior is rather common for low-dimensional magnets. It has been previously observed in Pb$_2$V$_3$O$_9$ (Refs.~\onlinecite{pb2v3o9,[{For example: }][{}]waki2004}), Cu(NO$_3)_2\cdot 2.5$H$_2$O (Ref.~\onlinecite{diederix-1978,*diederix-1978-2}), CuWO$_4$ (Ref.~\onlinecite{cuwo4-1981,*cuwo4-1991,*cuwo4-1996}), and other compounds. A distinctive feature of AgVOAsO$_4$ is the unusual regime of the interchain couplings. A close examination of the spin lattice shows that the interchain couplings $J_a$ and $J_c$ are frustrated. The four-bond unit depicted by the thick line in Fig.~\ref{fig:structure} comprises one FM and three AFM couplings, which can not be satisfied simultaneously (right panel of Fig.~\ref{fig:structure}).

Magnetic frustration has a dramatic effect on thermodynamic properties and ground state of spin-chain magnets. For example, frustrated interchain couplings induce an incommensurate magnetic structure in Cs$_2$CuCl$_4$ (Ref.~\onlinecite{coldea1997}) and a $\frac13$-magnetization plateau in Cs$_2$CuBr$_4$ (Ref.~\onlinecite{cs2cubr4}). Ising-like systems, such as Ca$_3$Co$_2$O$_6$ and Ca$_3$CoRhO$_6$, show an even more intricate behavior related to the formation of a partially ordered AFM state.\cite{kageyama1997,*sampathkumaran2004,*agrestini2008,niitaka2001,*sampathkumaran2002,*loewenhaupt2003} The frustrated nature of interchain couplings could similarly distinguish AgVOAsO$_4$ from other alternating-spin-chain magnets. The leading energy scale is, however, determined by the alternating $J$ and $J'$ bonds, and dictates the spin gap formation. Above the first critical field $H_c$, external magnetic field closes the spin gap, thereby a long-range magnetic order (BEC of magnons) is possible.\cite{bec-review,waki2004,diederix-1978,*diederix-1978-2} Since the frustration impedes the long-range order, one might expect the partial suppression of the BEC. The accessible critical fields ($\mu_0H_c\simeq 10$~T, $\mu_0H_s\simeq 48.5$~T) make AgVOAsO$_4$ an interesting compound for detailed high-field studies applying specific heat, magnetocaloric effect, ESR, and nuclear magnetic resonance measurements.


Turning to the structural aspects of the proposed spin model, we note that the frustration scenario in AgVOAsO$_4$ is very similar to one in M(VO)$_2$(PO$_4)_2$ with M = Ca and Sr.\cite{nath2008} In both cases, most of the couplings are long-range and AFM, yet the presence of the short-range FM coupling ($J_c$ in AgVOAsO$_4$) between the corner-sharing VO$_6$ octahedra makes the system frustrated. The magnitude of the frustration is controlled by the ratios of $J$'s. In M(VO)$_2$(PO$_4)_2$, four comparable couplings in the frustrated unit lead to the strong frustration on a 3D spin lattice, and the enhancement of quantum fluctuations manifests itself as the suppression of the specific heat maximum.\cite{nath2008} In AgVOAsO$_4$, $J$ and $J'$ dominate over $J_a$ and $J_c$, hence the magnetic behavior is quasi-1D, with the spin gap imposed by the alternating spin chains (similar to non-frustrated systems). The effects of the frustration should be primarily expected in high magnetic fields that close the spin gap and induce the long-range order.

\begin{figure}
\includegraphics{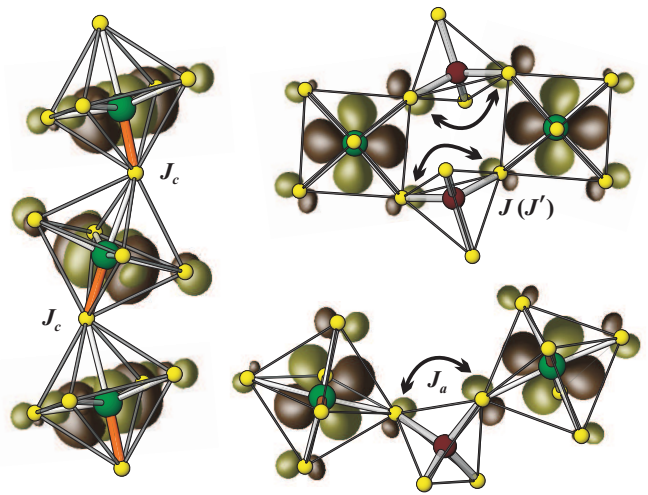}
\caption{\label{fig:wannier}
(Color online) Wannier functions for the superexchange pathways $J_c$ (left), $J$ and $J'$ (top right), and $J_a$ (bottom right). Thick (orange) lines denote short V--O(3) bonds.
}
\end{figure}

The peculiar superexchange regime of AgVOAsO$_4$ is readily understood from the examination of Wannier functions (Fig.~\ref{fig:wannier}). Owing to the symmetry of the half-filled vanadium orbital, the Wannier functions do not comprise the orbitals of the O(3) atom, and render the V--O--V superexchange pathway ($J_c$) inactive. Indeed, we find a weak FM exchange for $J_c$, despite the shortest V--V separation at this pathway. The leading exchanges $J$ and $J'$ run via longer pathways with double bridges of AsO$_4$ tetrahedra, since the edges of the tetrahedra provide a suitable overlap of the oxygen orbitals. Finally, the weaker AFM coupling $J_a$ can also be traced to a favorable overlap of the oxygen orbitals on the edge of the tetrahedron (a single bridge). All these findings conform to previous, both empirical\cite{roca1998} and microscopic\cite{tsirlin2010,bicu2po6} considerations regarding the superexchange in systems with tetrahedral polyanions: (i) the possible superexchange pathways are determined by the symmetry of the magnetic orbital; (ii) the tetrahedral anionic groups lead to short O--O distances and induce suitable overlaps of oxygen orbitals; (iii) double anionic bridges are more efficient than single bridges, owing to a different geometry and a larger number of O--O contacts. We expect that these simple arguments will facilitate further search for unusual spin lattices in V$^{+4}$ phosphates, arsenates, and other compounds with tetrahedral anions.

In summary, we have studied the crystal structure and magnetic properties of AgVOAsO$_4$, and proposed a microscopic description of this compound. This system basically conforms to the alternating spin chain model, although the quantitative understanding of the magnetic behavior requires an extension of the model toward a 3D spin lattice with frustrated couplings between alternating spin chains. Our magnetic susceptibility and high-field magnetization measurements establish a spin gap of about 13~K and a saturation field of 48.5~T. These estimates suggest the feasibility of further high-field measurements that should explore the influence of frustration on the behavior of dimer-based quantum magnets. The study of the Bose-Einstein condensation phenomena in AgVOAsO$_4$ might be particularly enlightening and calls for extensive experimental investigation of this compound.

\acknowledgments
We are grateful to Roman Shpanchenko, Yurii Prots, and Horst Borrmann for laboratory XRD measurements. Fruitful discussions with Alexander Smirnov are appreciated. We also acknowledge Franziska Weickert, Monika Gam\.za, Alexander Steppke, Manuel Brando, and Walter Schnelle for sharing our interest in AgVOAsO$_4$. We are grateful to ESRF for providing the beam time at ID31 and to Andy Fitch for his kind help during the data collection. A.T. appreciates the funding from Alexander von Humboldt foundation. Part of this work has been supported by EuroMagNET II under the EC contract 228043.


\begin{thebibliography}{58}
\bibitem [{not(){\natexlab{a}}}]{note1}%
  \BibitemOpen
  \bibinfo {note} {The refinement at 20~K led to a somewhat higher $R_p=0.090$
  compared to $R_p=0.068$ at 300~K. This difference, as well as the slightly
  underestimated ADP of oxygen, could be related to strains that develop at low
  temperatures and induce an anisotropic reflection broadening.}\BibitemShut
  {Stop}%
\bibitem [{not(){\natexlab{b}}}]{note2}%
  \BibitemOpen
  \bibinfo {note} {At 300~K, we even had to refine the ADP of silver in
  anisotropic approximation with $U_{11}=0.0069$~\r A$^2$,
  $U_{22}=0.0034(2)$~\r A$^2$, $U_{33}=0.0281(4)$~\r A$^2$,
  $U_{12}=0.0006(4)$~\r A$^2$, $U_{13}=0.0014$~\r A$^2$, and
  $U_{23}=-0.0021(3)$~\r A$^2$.}\BibitemShut {Stop}%
\bibitem [{not(){\natexlab{c}}}]{note5}%
  \BibitemOpen
  \bibinfo {note} {We varied the lower boundary of the fitting range in order
  to obtain stable values for the fitting parameters $C,\theta$, and~$\chi_0$.}\BibitemShut {Stop}%
\bibitem [{not(){\natexlab{d}}}]{note3}%
  \BibitemOpen
  \bibinfo {note} {Here, we use the powder-averaged $\bar g=1.947$ from
  ESR.}\BibitemShut {Stop}%
\end{thebibliography}
%
\newpage
\ 
\ 
\newpage
\renewcommand{\thefigure}{S\arabic{figure}}
\setcounter{figure}{0}
\begin{widetext}
\begin{center}
 \large
 \centerline{\textbf{Supplementary material}}
\end{center}
\bigskip

\begin{figure}[!ht]
\includegraphics[width=10cm]{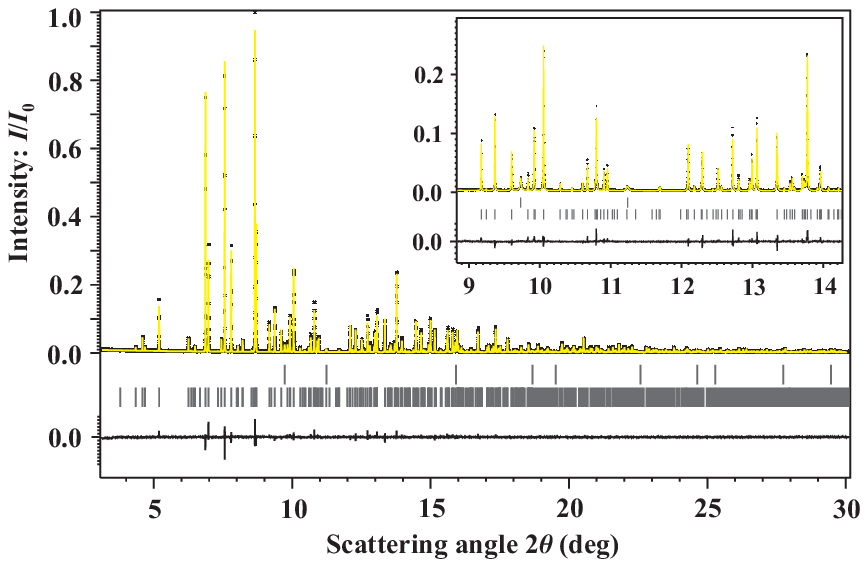}
\caption{\label{fig:s1}
Crystal structure refinement at 300~K: experimental (symbols), calculated (yellow line), and difference (black line) patterns. Ticks show the reflections of AgVOAsO$_4$ (lower line) and the Ag impurity (upper line).
}
\end{figure}
\bigskip

\begin{figure}[!ht]
\includegraphics[width=10cm]{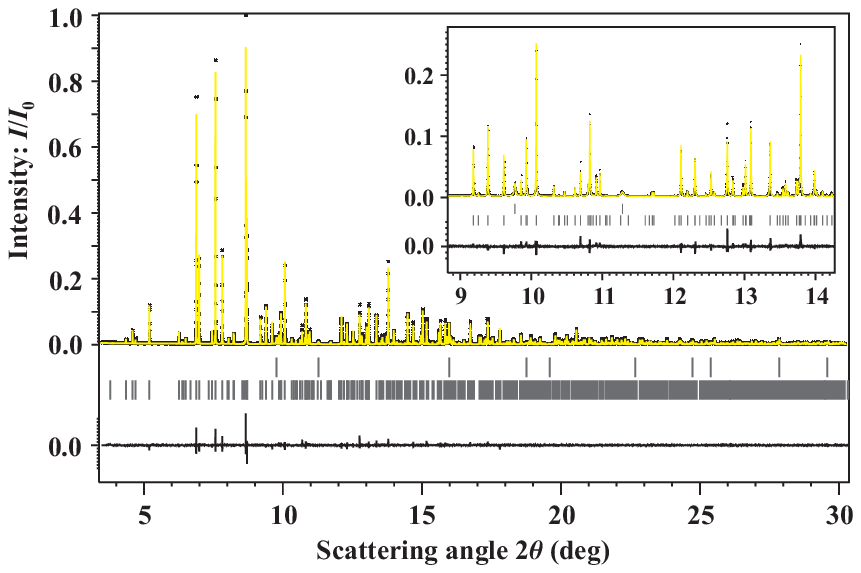}
\caption{\label{fig:s2}
Same as Fig.~\ref{fig:s1} for the structure refinement at 20~K.
}
\end{figure}
\bigskip

\begin{figure}[!ht]
\includegraphics[width=10cm]{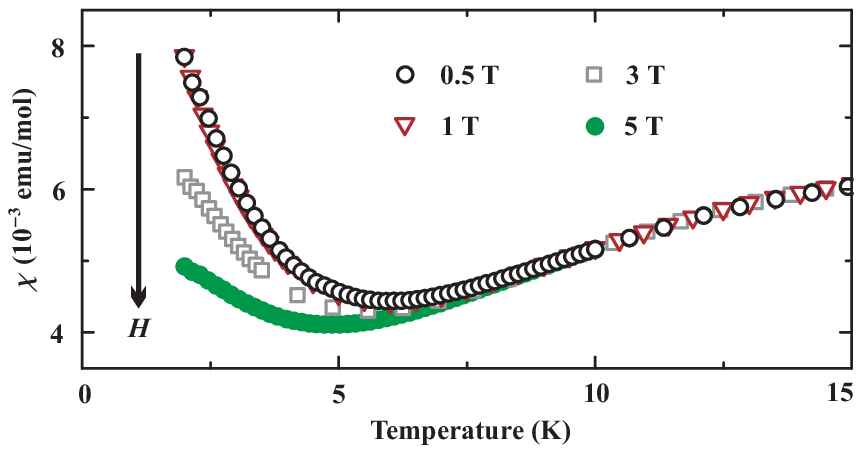}
\caption{\label{fig:s3}
Magnetic susceptibility of AgVOAsO$_4$ measured in the applied fields of 0.5, 1, 3, and 5~T. The arrow shows the reduction in the low-temperature upturn upon the increase in the magnetic field $H$.
}
\end{figure}

\begin{figure}[!ht]
\includegraphics[width=10cm]{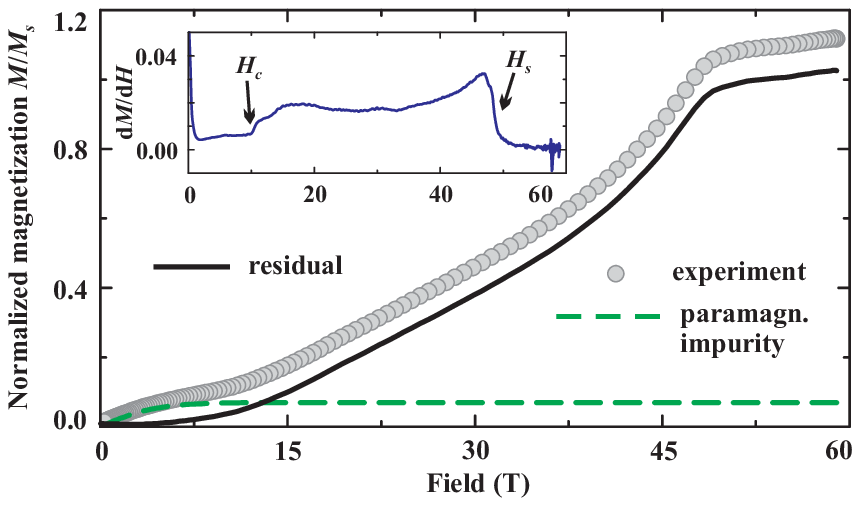}
\caption{\label{fig:s4}
Experimental high-field magnetization curve measured at 1.4~K (circles), the contribution of the paramagnetic impurity (dashed line), and the intrinsic magnetization (solid line) of AgVOAsO$_4$ obtained by the subtraction. The inset shows the field derivative of the experimental magnetization curve along with the critical fields $H_c$ (closing of the spin gap) and $H_s$ (saturation).
}
\end{figure}
\end{widetext}

\end{document}